\documentclass[aps,prb,twocolumn,superscriptaddress,showpacs,10pt]{revtex4-1} 

\usepackage{amssymb}
\usepackage{graphicx}
\usepackage{amsmath}
\usepackage{dsfont}
\usepackage{bbm}
\begin{document}

\title{From Andreev bound states to Majorana fermions in topological wires on superconducting substrates : a story of  mutation}
\author{D. Chevallier}
\email{denis.chevallier@gmail.com}
\affiliation{Laboratoire de Physique des Solides, CNRS UMR-8502, Universit\'e Paris Sud, 91405 Orsay Cedex, France}
\affiliation{Institute de Physique Th\'eorique, CEA/Saclay, Orme des Merisiers, 91190 Gif-sur-Yvette Cedex, France}
\author{P. Simon}
\affiliation{Laboratoire de Physique des Solides, CNRS UMR-8502, Universit\'e Paris Sud, 91405 Orsay Cedex, France}
\author{C. Bena}
\affiliation{Laboratoire de Physique des Solides, CNRS UMR-8502, Universit\'e Paris Sud, 91405 Orsay Cedex, France}
\affiliation{Institute de Physique Th\'eorique, CEA/Saclay, Orme des Merisiers, 91190 Gif-sur-Yvette Cedex, France}

\date{\today}

\begin{abstract}
We study the proximity effect in a topological nanowire tunnel coupled to an s-wave superconducting substrate.
We use a general Green's function approach that allows us to study the evolution of the Andreev bound states in the wire into Majorana fermions. We show that the strength of the tunnel coupling induces a topological transition in which the Majorana fermionic states can be destroyed when the coupling is very strong.  Moreover, we provide a phenomenologial study of the effects of disorder in the superconductor on the formation of Majorana fermions. We note a non-trivial effect of a quasiparticle broadening term which can take the wire from a topological into a non-topological phase in certain ranges of parameters. Our results have also direct consequences for a nanowire coupled to an inhomogenous superconductor.   
\end{abstract}


\maketitle

\section{Introduction}\label{introduction}

Majorana fermionic states have received a lot of interest because of 
their exotic properties \cite{beenakker, alicea} such as non-Abelian statistics, that open the perspective of using them for quantum computation \cite{nayak_dassarma,alicea_fisher}. A system which is expected to exhibit such states consists of a semiconducting wire such as InAs or InSb \cite{spin_orbit_nanowire}, for which the spin-orbit coupling is strong, in the presence of an applied Zeeman field and in the proximity of an s-wave superconductor (SC) \cite{lutchyn_dassarma,oreg_vonoppen}.  Moreover, a few recent experiments have reported the observation of zero-bias peaks in such systems, \cite{kouwenhoven, heiblum, caroff} which are in good agreement but not completely consistent with the existence of these Majorana fermions. Such experimental observations have been shown to also have simpler explanations because a zero bias peak could also arise without any Majorana fermions \cite{potter_lee, atland,pientka,klinovaja2, aguado_defranseschi, marcus, lee_defranseschi, finck_li}. 

Many theoretical studies have studied the properties of the topological wire when in contact with an homogeneous SC substrate \cite{loss11,stoudenmire,bena_sticlet,lutchyn_fisher,cool_franz,stanescu_dassarma,aguado,klinovaja1,tanaka}. However, the effect of an inhomogenous SC on the Majorana states has received less attention \cite{meyer_refael, lobos_dassarma, trauzettel_nagasoa, klapwijk, sau_dassarma, kells_brouwer, prada_aguado, stanescu_tewari}.
In a previous work, we have also studied the Andreev bound states (ABSs) which arise in such a wire when it is end contacted with superconductors. These states, which are a reminiscence of the quantized modes of the nanowire in the normal state \cite{bena} can mutate into Majorana Fermions (MFs) \cite{chevallier_mutation,chevallier_long}. 

In this article we make use of a more universal approach allowing one to characterize quite generally the proximity of various superconducting substrates. In this approach the proximity effect is induced by a local hopping term between the wire and the SC substrate. We analyze the differences between the predictions of this model and those of the more widely used one in which the proximity SC gap induced in the wire is added by hand as a constant pairing term. This general formalism allows us to describe the ABSs which arise in the topological wire in the proximity of the SC substrate, as well as their evolution into MFs when the spin-orbit coupling and the Zeeman field are turned on. Moreover, other properties of the superconducting substrate such as disorder, can be easily take into account in this approach. Thus, we apply our model to an s-wave superconductor with a finite broadening of the SC coherence peaks, consistent with a finite quasiparticle lifetime \cite{dynes}. We study the effect of this finite lifetime, as well as of the tunneling rate between the superconductor and the wire on the formation of Majorana states. We show that increasing the tunnel coupling induces a topological transition which splits the zero-energy Majorana peak; interestingly enough, this peak can in certain conditions reform by increasing the quasiparticle broadening, but this happens only for a very finely tuned regime of parameters, and it is unlikely to have observable experimental consequences.  We also note that, as expected, with increasing the quasiparticle broadening the Majorana peaks widen and get dissolved in the bulk. 

The paper is organized as follows. In Sec. II we present the model of a semiconducting nanowire connected via a tunnel coupling with an s-wave superconducting substrate. In Sec. III we describe the formation of the MFs from the ABSs in such a system. Finally, in Sec. IV, we focus on the possibility of generalizing this Green's function formalism to various types of s-wave superconducting substrates, and we illustrate it by studying the effects of a finite SC quasiparticle broadening on the Majorana physics. We conclude in Sec. V.

\section{Model}\label{section}

\begin{figure}[ht]
	\centering
		\includegraphics[width=7cm]{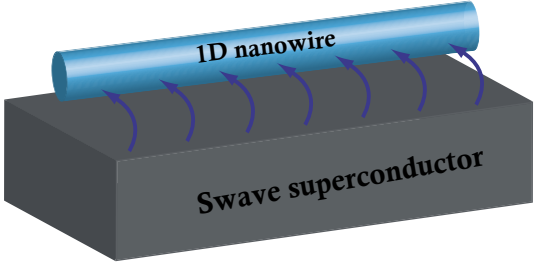}
	\caption{(Color online) Scheme of our system: a semiconducting nanowire (Bare Green's function $\tilde{G}_0(\omega)$) connected via a tunnel coupling to a superconducting substrate (Superconducting self energy $\tilde{\Sigma}_S(\omega)$).}
	\label{fig:setup}
\end{figure}

The formalism that we will be using allows one to describe the physics of a semiconducting nanowire in contact with any type of superconducting substrate. We focus on a wire with strong spin-orbit coupling, in the presence of a Zeeman field (see Fig. \ref{fig:setup}). The total Hamiltonian for such a system has the form
\begin{equation}
H= H_{NW} + H_{S} + H_{T}.
\label{hamilt_total}
\end{equation}

In terms of the extended Nambu spinors the creation operator of the nanowire electrons is given by $\hat{c}^\dagger = ( c_{\uparrow}^\dagger, c_{\downarrow}^\dagger, c_{\downarrow}, -c_{\uparrow} )$, and the semiconducting nanowire Hamiltonian becomes
\begin{equation}
H_{NW} = \int \hat{c}^\dagger\left[ (\frac{p^2}{2m}-\mu)\tau_z+\alpha p \sigma_y \tau_z + V_z\sigma_z\right]\hat{c}\:dx,
\label{hamilt_nw}
\end{equation}
where $\sigma$ and $\tau$ are the Pauli matrices respectively in spin and particle-hole spaces, $\mu$ is the chemical potential, $\alpha$ is the strength of the Rashba spin-orbit coupling and $V_z$ is the applied Zeeman magnetic field. The Hamiltonian of the bulk superconductor can be written as
\begin{equation}\label{hamilt_superconduc}
	H_{S}=\sum_{\mathbf k,\sigma}\xi_{k} \Psi^{\dagger}_{\mathbf k,\sigma}\Psi_{\mathbf k,\sigma}+\Delta \Psi^{\dagger}_{\mathbf k,\uparrow}\Psi^{\dagger}_{-\mathbf k,\downarrow}+h.c.	
\end{equation}
with $\xi_{k}=\frac{k^{2}}{2m}-\mu$ and $\Psi_{\mathbf k,\sigma}$ is the annihilation operator for an electron in the superconductor, having spin $\sigma$ and momentum $\mathbf k$. The hopping term between the SC and the nanowire takes the following form
\begin{equation}
	H_{T}=\sum_{j,\sigma}  \tilde{t}_j c^{\dagger}_{j,\sigma} \sum_{\mathbf k} \Psi^{\dagger}_{\mathbf k,\sigma} e^{i k_x j} + h.c.
\end{equation}
where the operator $c^{\dagger}_{j,\sigma}$ creates an electron with spin $\sigma$ on a site labeled by $j$. Since the total Hamiltonian is quadratic in the SC degrees of freedom, we can integrate out these modes, such that the effect of the SC substrate is taken into account by dressing the bare Green's function of the nanowire  by a superconducting self-energy 
$
\tilde{G}_R^{-1} (\omega) = \tilde{G}^{-1}_{0 R} (\omega)-\tilde{\Sigma}_R^{S}(\omega).
$
The total retarded superconducting self-energy can be written as $\tilde{\Sigma}_R^{S}(\omega)={\cal I}\otimes\tilde{\Sigma}_{j,R}^{S}(\omega)$, with ${\cal I}$ being the unity matrix in the space of sites, and $\tilde{\Sigma}_{j,R}^{S}$ the on-site retarded superconducting self-energy given by
\begin{align}
\tilde{\Sigma}_{j,R}^{S} (\omega) =\left|\tilde{t}_j\right|^2 \tau_z~\tilde{g}_R(\omega)~ \tau_z.
\label{eq-TunnSelfE}
\end{align}

The local self energy depends on the tunneling amplitude $\tilde{t}_j$ as well as on the retarded Green's function for the superconducting bulk electrons and can be written as \cite{levy_yeyati, chevallier_double_dot}
\begin{equation}\label{supercond_green}
\hat{\Sigma}_{j,R}^S(\omega)=\Gamma_{j,S}\tilde{g}_R(\omega)=\Gamma_{j,S} \frac{\omega}{\sqrt{\Delta^2-\omega^2}}\left({\bf 1}+\frac{\Delta}{\omega} \tau_x\right)
\end{equation}
where {\bf 1} is the unity matrix in spin/particle-hole space, and $\Gamma_{j,S}=\pi\nu(0)\left|\tilde{t}_j\right|^2$ is the tunneling rate. We take this to be uniform along the nanowire for most of the rest of the paper,  $\Gamma_{j,S}=\Gamma_S$, $\tilde{t}_j=\tilde{t}$. 

The retarded bare Green's function of the semiconducting nanowire electrons \cite{jonckheere} is given by
\begin{equation}\label{bare_green}
\tilde{G}_{0 R}^{-1}(\omega)=(\omega+i \delta){\bf 1}-H_{NW}
\end{equation}
where $\delta$ is an infinitesimal quasiparticle inverse lifetime which is introduced to avoid divergences in the numerical evaluations. 
For the purpose of our analysis, the Hamiltonian of the nanowire is best described using a lattice tight-binding model, such that Eq. (\ref{hamilt_nw}) becomes
\begin{align}\label{hamilt_nw_discret}
H_{NW}=\sum^N_j &c^\dagger_j \left[(t-\mu)\tau_z+V_z \sigma_z\right]c_j \notag\\
&-\frac{1}{2}c^\dagger_j\left[t\tau_z+i\alpha\sigma_y\tau_z + h.c.\right]c_{j+1} ,
\end{align}
where we defined the creation operator $c^\dagger_j$ of an electron in the nanowire on site $j$
in the Nambu basis as $c^\dagger_j=( c_{j\uparrow}^\dagger, c_{j\downarrow}^\dagger, c_{j\downarrow}, -c_{j\uparrow} )$. Here and throughout the remainder of the text, we work with physical dimensions corresponding to $\hbar=1$, and in units in which the hopping term between sites is $t=1$. In our calculations we focus on obtaining the local density of states (LDOS) (evaluated on a given site $j$) in the wire,  which is given by the imaginary part of the dressed Green's function
\begin{equation}\label{eq_dos}
n_j(\omega)=-\frac{1}{\pi}\sum_{\beta=\uparrow,\downarrow}\textrm{Im}[\tilde{G}^{jj,\beta\beta}_{R}(\omega)].
\end{equation}

Throughout this paper we focus on the limit where the system is in the topological phase by chosing $V_z=0.4, \Delta=0.3, \mu=0, \textrm{and}\; \alpha=0.2$. A finite width ($\delta=0.002 \hbar v_F/a$) is introduced in the numerical evaluations yielding a finite width of the peaks in the LDOS. We consider also that we are in the regime $l\ll\xi$ where $\xi$ is the superconducting coherence length and $l$ the length of the nanowire.

\section{From Andreev Bound States to Majorana Fermions}\label{formation_majorana}

\begin{figure}[ht]
	\centering
		\includegraphics[width=6cm]{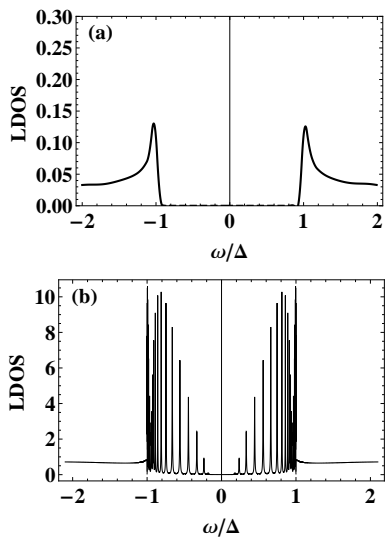}
	\caption{LDOS of two models:  DOS dependence on energy when the superconductivity is put ``by hand'' (a) and when the superconductivity is induced via a tunnel coupling (b).}
	\label{fig:ind_gap}
\end{figure}

 We focus first on the qualitative features of the proximity effect induced by the tunnel coupling with the SC substrate, and on the differences with the model in which the pairing term is put ``by hand'' directly in the nanowire \cite{lutchyn_dassarma,oreg_vonoppen,chevallier_mutation,chevallier_long}. As described in Fig. \ref{fig:ind_gap}a, in the latter case a superconducting gap $\Delta$ is induced in the nanowire (in absence of Zeeman field and Rashba spin-orbit coupling). When we turn on the Zeeman field the effective gap becomes $\Delta-V_z$, and for $V_z$ equal to $\Delta$ the gap closes. For values of $V_z$ larger than $\Delta$ and in the presence of spin-orbit coupling, the gap reopens and is topological in nature, and two Majorana states form at zero energy. The Rashba spin-orbit coupling localizes the two Majorana states more and more at the two ends of the nanowire. 

When the proximity effect is modeled via a hopping term, in the absence of a Zeeman field and of Rashba spin-orbit coupling, a gap-like feature also appears at an energy $\Delta$, which can be described as a SC pseudo-gap (see Fig. \ref{fig:ind_gap}b ). However, inside this gap we note the formation of ABSs, the number of these states being determined by the ratio between the inverse nanowire length $1/l$ and the size of the gap (i.e. the energy difference between two ABSs is proportional to $1/l$)\cite{bena}.  

\begin{figure*}[ht]
	\centering
		\includegraphics[width=14cm,height=13cm]{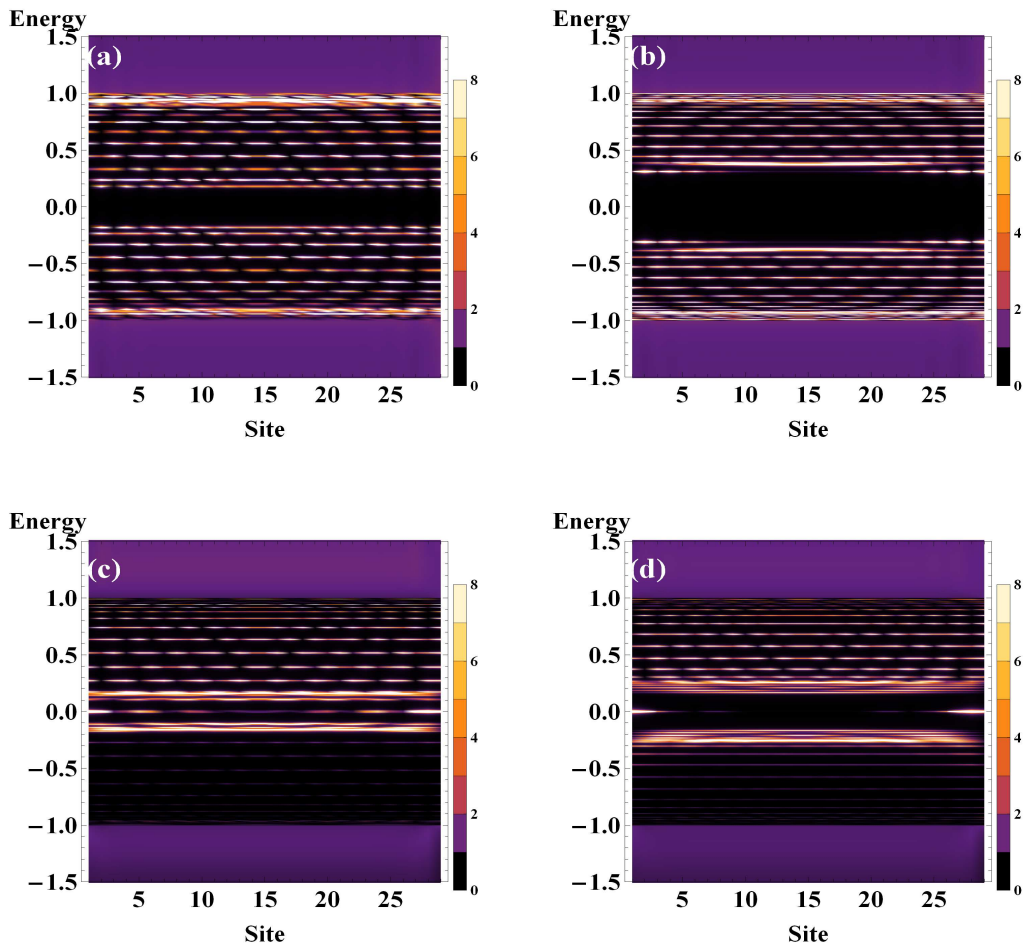}
	\caption{(Color online) LDOS as a function of energy (in units of $\Delta$) and position for $\Delta=\Gamma_S=0.3$ and for various values of $\alpha$ and $V_z$. a) non-topological state, no Zeeman field and no spin-orbit coupling; b) non-topological state with a finite spin-orbit coupling ($\alpha=0.1$); c) topological state with a not too large spin-orbit coupling ($V_z=0.4$, $\alpha=0.1$); d) topological state with strong spin-orbit coupling ($V_z=0.4$, $\alpha=0.3$).}
	\label{ldos-abs}
\end{figure*}

In Fig. \ref{ldos-abs} we have plotted the LDOS as a function of energy and position for various parameters in order to characterize the evolution of ABS. In the absence of spin-orbit coupling and magnetic field, these ABSs have a uniform weight along the wire (see Fig.~\ref{ldos-abs}a ) and are a reminiscence of the quantized modes of the wire in the normal state \cite{bena} (note that Fig. \ref{fig:ind_gap}b corresponds to an energy cut of Fig. \ref{ldos-abs}a  at a given position).  
Turning on the spin-orbit coupling localizes these states towards the ends of the wire (see Fig.~\ref{ldos-abs}b ), even in the non-topological state. Including a Zeeman field breaks the spin degeneracy and splits each of these states. The two ABSs come closer and closer with increasing the magnetic field until when, for a certain value of $V_z$, and in the presence of spin-orbit coupling, they merge (note this merging is also controlled by the chemical potential) (see Fig.~\ref{ldos-abs}c ) and form two Majorana states. Increasing the spin-orbit coupling localizes these states more and more towards the ends of the wire (see Fig.~\ref{ldos-abs}d ).

Thus we see that the proximity effect induced via the tunnel coupling between the SC and the nanowire also allows for the formation of localized Majorana modes, which can be seen evolving from the Andreev bound states in the wire. 

In order to understand how these Majorana states are formed in a more quantitative manner, we analyze the effective Hamiltonian induced in the wire via the coupling with the superconducting substrate. Thus, we note that if we focus on the low-energy sector where we expect the Majorana modes to form, the self-energy induced in the wire  is equivalent with an effective SC gap which can be evaluated by setting the energy to zero in the second term in Eq.~(\ref{supercond_green}), yielding $\Delta_{\rm eff}=\Gamma_S$.  It is this effective gap which governs the transition to a non-topological phase when the Zeeman field becomes smaller than a critical value, $V_z^2<\Delta_{\rm eff}^2+\mu^2$.

\begin{figure}[ht]
	\centering
		\includegraphics[width=6cm]{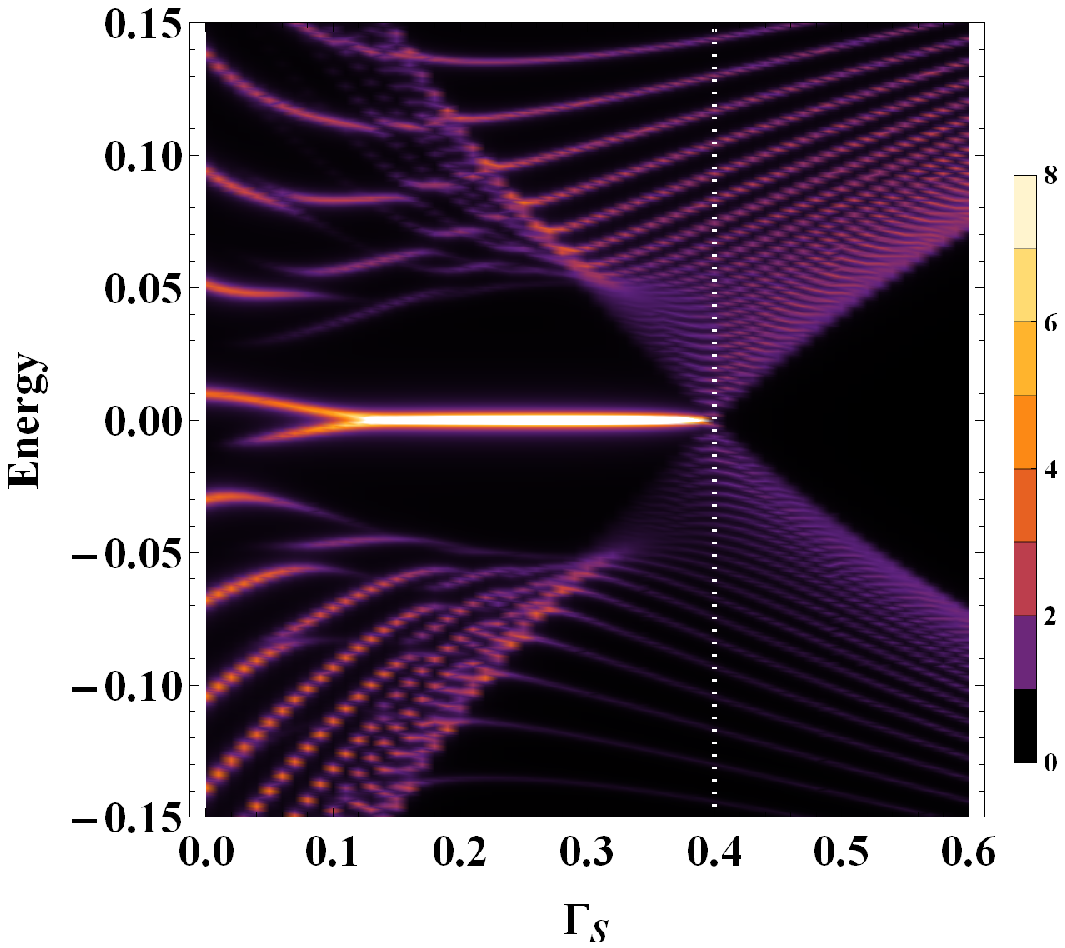}
	\caption{(Color online) LDOS of a Majorana peak (on the first site of the wire) as a function of energy  and the tunneling rate $\Gamma_S$. The dashed line denotes the transition between a topological and non-topological state which occurs at $\tilde{\Gamma}_S\approx0.4$ (with our parameters).}
	\label{fig:tran_amprate0}
\end{figure}

Alternatively seen, for large values of $\Delta_{\rm eff}$ a transition to a non-topological phase occurs. In terms of the tunnel coupling between the SC and the wire the corresponding condition to exit the topological phase is given by $\Gamma_S >\tilde{\Gamma}_S$, where $\tilde{\Gamma}_S=\sqrt{V_z^2-\mu^2}$. This can be checked numerically, and in Fig. \ref{fig:tran_amprate0} we have plotted the LDOS at one end of the nanowire as a function of the energy and $\Gamma_S$. Indeed, we note that for values of $\Gamma_S$ larger than $\tilde{\Gamma}_S$ the Majorana peaks split in two, marking the exit from the topological phase. We have checked that the energy difference $\delta E$ between these ABSs depends linearly on $\Gamma_S$.

We also note that for very small tunnel coupling parameters the Majorana states are also split; this behavior can be explained by the fact that  in this regime the superconductivity has not sufficiently penetrated the nanowire to have a real superconducting gap. By increasing the nanowire length the splitting of the Majorana states occurs for smaller and smaller values of $\Gamma_S$, confirming that this is a finite size effect; for large nanowire lengths we obtain a robust zero bias peak even for very small transmission rates. 

While we do not detail it here, this formalism allows one to consider also a non-uniform tunneling rate which corresponds to disorder at the interface between the wire and the superconducting substrate. Such disorder yields a position dependent effective superconducting gap in the wire, and the Majorana modes are destroyed for large disorder strengths.

\section{Disordered superconducting substrate}

This general formalism allows us to model also a disordered as well as an inhomogenous  superconductor. Such properties can be encoded in the SC Green's function via the superconducting self energy. We first use our model to study the effect of a finite quasiparticle broadening which is phenomenologically taken into account via a non-zero energy imaginary part  $\gamma$, $\omega \rightarrow \omega+i \gamma$ in Eq. (\ref{supercond_green}) \cite{dynes}. This broadening models a finite quasiparticle lifetime, and is characterized by a smoothed BCS gap at $\omega=\Delta$ and by the presence of states inside the SC gap.

\begin{figure}[ht]
	\centering
		\includegraphics[width=6cm]{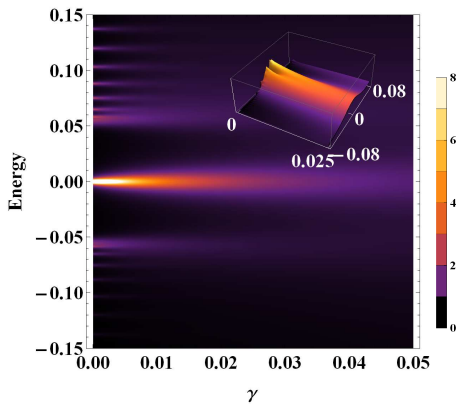}
	\caption{(Color online) LDOS of a Majorana peak (on the first site of the wire) as a function of the energy and the inverse quasiparticle lifetime $\gamma$ . We take $\Gamma_S=0.3$ in the main plot (no topological transition with varying $\gamma$) and in the inset we take $\Delta=0.15$, $V_z=1$,  and $\Gamma_S=1.01$ (a Majorana peak is expected to form when $\gamma>\tilde{\gamma}\approx0.021$). }
	\label{fig:tran_amprate}
\end{figure}

In terms of the broadening parameter $\gamma$, the new effective gap becomes $\Delta_{eff}=\Gamma_S\Delta/\sqrt{\Delta^2+\gamma^2}$, as it can be seen from Eq. (\ref{supercond_green}), and the condition to have a topological phase is given by $\gamma>\tilde{\gamma}=\Delta\sqrt{\frac{\Gamma^2_S}{V_z^2-\mu^2}-1}$. Thus, the value of $\gamma$ for which the transition takes place depends essentially on the difference $\Gamma_S^2-V_z^2$. For large magnetic fields $V_z>\Gamma_S$ the topological condition is satisfied for all values of $\gamma$, and no transition to a non-topological state is possible. In Fig. \ref{fig:tran_amprate} we have plotted the LDOS of a Majorana peak (i.e. the LDOS on the first site of the wire) as a function of energy and $\gamma$ in this regime. Note that indeed we cannot identify any topological transition, however the Majorana peaks widen and get dissolved in the bulk when $\gamma$ becomes of the same order of magnitude as the SC gap. This is not very surprising as for such strong pair-breaking the BCS DOS is smoothed out and becomes similar to that of a normal metal, and our system behaves as a topological wire in the non-SC (``normal'') regime (no Majorana modes).

When $\Gamma_S>V_z$  the Majorana fermions do not form for $\gamma=0$, however, for certain parameter values chosen such that $\tilde{\gamma}$ is small with respect to the topological gap ($\Gamma_S$ very close to $\tilde{\Gamma}_S$, large $V_z$ with respect to $\Delta$) there exists a transition to a topological phase with increasing $\gamma$. Finding the range of parameters in which this can be realized requires fine tuning, and in the inset in Fig. \ref{fig:tran_amprate} we have illustrated such a situation ($\Delta=0.15$, $V_z=1$, $\Gamma_S=1.01$), for which a phase transition exists and the Majorana peak reforms for $\gamma>{\gamma}\approx0.021$

We focus also on an inhomogenous superconductor consisting of regions with different tunnel couplings between the wire and the SC, as well as with different quasiparticle broadenings. In particular we consider a superconductor containing three regions, a central region in the topological phase ($\Gamma_S<\tilde{\Gamma}_S$, $\gamma=0$), and two exteriors ones for which the quasiparticle broadening as well as the tunnel coupling can be tuned such that the regions are in either the topological or the non-topological phase. When all regions are in topological phase, we restore a well-known case where we only have two Majorana states localized at the ends of the nanowire. When only the central region is in the topological phase, two Majorana fermions form at the extremities of the central part.  An interesting situation is that of two exteriors regions with very strong broadening. In this kind of configuration, the exteriors regions are quasi-normal, and we recover the behavior described in Ref. ~\onlinecite{chevallier_mutation} for SN junctions, i.e.,  two Majorana fermionic states extended over the entire "normal" parts, as well as over a broad energy range. 
 
Our approach can also be applied to other kinds of disorder such as magnetic disorder. Indeed, magnetic impurities strongly modify the excitations spectrum of the superconductor, as it has been shown for the first time by Abrikosov and Gorkov \cite{abrikosov}. In principle, this type of disorder can be treated by solving the Usadel equations in order to obtain the associated Green's function of the superconductor in presence of a finite density of magnetic impurities\cite{usadel,houzet}.  The resulting Green's functions need then to be injected in our formalism (see Eq.~(\ref{supercond_green}) to recover an effective local self-energy in the wire, and allow one to test for the formation and destruction of Majorana fermions as a function of disorder. However, solving such equations is a complex task, involving both numerical and analytical techniques, which we believe is a research subject in itself, and well beyond the scope of the present work; this subject will be addressed elsewhere \cite{usadel-mi}.

\section{Conclusion}\label{conclusion}

To summarize, we have used a general approach to describe the proximity effect induced in a semiconducting nanowire connected to a superconducting substrate via a tunnel coupling. We have described the evolution of the ABSs into MFs when turning on the spin-orbit coupling and the Zeeman field in the wire. We have also shown that the nanowire can reach a non-topological phase for strong values of the tunneling rate between the SC and the wire. Moreover, we have applied our method to study the effects of the disorder in the superconductor on the physics of the Majorana fermions in the wire. In particular we have focused on the effects of quasiparticle broadening, and we have found that a finite quasiparticle lifetime can also take the system through a topological phase transition in a finely tuned range of parameters. Finally, we have shown that Majorana states can form in an inhomogenous superconductor.  
\acknowledgments

D. C. and C. B. acknowledge discussions with T. Jonckheere and L. P. Kouwenhoven. The work of C. B. and D. C. is supported by the ERC Starting Independent Researcher Grant NANO-GRAPHENE 256965.


\begin{thebibliography}{99}

\bibitem{beenakker} C. W. J. Beenakker,  Annu. Rev. Con. Mat. Phys. {\bf 4}, 113 (2013).
 
\bibitem{alicea} J. Alicea, Rep. Prog. Phys. {\bf 75}, 076501 (2012).

\bibitem{nayak_dassarma} C. Nayak, S. H. Simon, A. Stern, M. Freedman, and S. Das Sarma, Rev. Mod. Phys. {\bf 80}, 1083 (2008).

\bibitem{alicea_fisher} J. Alicea, Y. Oreg, G. Refael, F. von Oppen, and M. P. A. Fisher, Nat. Phys. {\bf 7}, 412 (2011).

\bibitem{spin_orbit_nanowire} S. Nadj-Perge, S. M. Frolov, E. P. A. M. Bakkers, and L. P. Kouwenhoven, Nature {\bf 468}, 1084 (2010); H. A. Nilsson {\it et al.},
Nano Letters {\bf 9}, 3151 (2009); V. Aleshkin {\it et al.}, Semiconductors {\bf 42}, 828 (2008), ISSN 1063-7826. 

\bibitem{lutchyn_dassarma} R. M. Lutchyn, J. D. Sau, and S. Das Sarma, Phys. Rev. Lett. {\bf 105}, 077001 (2010).

\bibitem{oreg_vonoppen} Y. Oreg, G. Refael, and F. von Oppen, Phys. Rev. Lett. {\bf 105}, 177002 (2010). 

\bibitem{kouwenhoven} V. Mourik, K. Zuo, S. M. Frolov, S. R. Plissard, E. P. A. M. Bakkers, and L. P. Kouwenhoven, Science {\bf 336}, 1003 (2012).

\bibitem{caroff} M. T. Deng, C. L. Yu, G. Y. Huang, M. Larsson P. Caroff, H. Q. Xu, Nano Lett. {\bf 12}, 6414 (2012).

\bibitem{heiblum} A. Das, Y. Ronen, Y. Most, Y. Oreg, M. Heiblum, H. Shtrikman, Nat. Phys. {\bf 8}, 887 (2012).

\bibitem{potter_lee} J. Liu, A. C. Potter, K. T Law, P. A Lee, Phys. Rev. Lett. {\bf 109}, 267002 (2012).

\bibitem{atland} D. Bagrets and A. Altland, Phys. Rev. Lett. {\bf 109}, 227005 (2012).

\bibitem{pientka} F. Pientka, G. Kells, A. Romito, P. W. Brouwer, and F. von Oppen, Phys. Rev. Lett. {\bf 109}, 227006 (2012).

\bibitem{klinovaja2} D. Rainis, L. Trifunovic, J. Klinovaja, and D. Loss, Phys. Rev. B {\bf 87}, 024515 (2013).

\bibitem{aguado_defranseschi} E. J. H. Lee, X. Jiang, R. Aguado, G. Katsaros, C. M. Lieber, S. De Franceschi, Phys. Rev. Lett. {\bf 109}, 186802 (2012).

\bibitem{marcus} H. O. H. Churchill, V. Fatemi, K. Grove-Rasmussen, M. T. Deng, P. Caroff, H. Q. Xu, and C. M. Marcus, Phys. Rev. B {\bf 87}, 241401(R) (2013).

\bibitem{lee_defranseschi} E. J. H. Lee, X. Jiang, M. Houzet, R. Aguado, C. M. Lieber, S. De Franceschi, arXiv:1302.2611.

\bibitem{finck_li} A. D. K. Finck, D. J. Van Harlingen, P. K. Mohseni, K. Jung, and X. Li, Phys. Rev. Lett. {\bf 110}, 126406 (2013). 

\bibitem{loss11} S. Gangadharaiah, B. Braunecker, P. Simon, and D. Loss, Phys. Rev. Lett. {\bf 107}, 036801 (2011). 

\bibitem{stoudenmire} E. M. Stoudenmire, J. Alicea, O. A. Starykh, and M. P.A. Fisher, Phys. Rev. B {\bf 84}, 014503 (2011).

\bibitem{bena_sticlet} D. Sticlet, C. Bena, and P. Simon, Phys. Rev. Lett. {\bf 108}, 096802 (2012).

\bibitem{lutchyn_fisher} R. M. Lutchyn, and M. P. A. Fisher, Phys. Rev. B {\bf 84}, 214528 (2011). 

\bibitem{cool_franz} A. Cook, and M. Franz, Phys. Rev. B {\bf 84}, 201105 (2011).

\bibitem{stanescu_dassarma} T. Stanescu, R. M. Lutchyn, S. Das Sarma, Phys. Rev. B {\bf 84}, 144522 (2011). 

\bibitem{aguado} P. San-Jose, E. Prada, and R. Aguado, Phys. Rev. Lett. {\bf 108}, 257001 (2012).

\bibitem{klinovaja1} J. Klinovaja and D. Loss, Phys. Rev. B {\bf 86}, 085408 (2012).

\bibitem{tanaka} Y. Asano and Y. Tanaka, Phys. Rev. B {\bf 87}, 104513 (2013).

\bibitem{meyer_refael} J. S. Meyer and G. Refael, Phys. Rev. B {\bf 87}, 104202 (2013).

\bibitem{lobos_dassarma} S. Takei, B. M. Fregoso, H-Y. Hui, A. M. Lobos, S. Das Sarma, Phys. Rev. Lett. {\bf 110}, 186803 (2013).

\bibitem{trauzettel_nagasoa} S. Nakosai, J. C. Budich, Y. Tanaka, B. Trauzettel, N. Nagaosa, Phys. Rev. Lett. {\bf 110}, 117002 (2013).

\bibitem{klapwijk} E. F. C. Driessen, P. C. J. J. Coumou, R. R. Tromp, P. J. de Visser, T. M. Klapwijk, Phys. Rev. Lett. {\bf 109}, 107003 (2012).

\bibitem{sau_dassarma} J. D. Sau, S. Das Sarma, Phys. Rev. B {\bf 88}, 064506 (2013).

\bibitem{kells_brouwer} G. Kells, D. Meidan, and P. W. Brouwer, Phys. Rev. B {\bf 86},100503(R) (2012).

\bibitem{prada_aguado} E. Prada, P. San-Jose, and R. Aguado, Phys. Rev. B {\bf 86}, 180503(R) (2012).

\bibitem{stanescu_tewari} T. D. Stanescu and S. Tewari, Phys. Rev. B {\bf 87}, 140504(R) (2013).    

\bibitem{bena} C. Bena, Eur. Phys. J. B {\bf 85}, 196 (2012).

\bibitem{chevallier_mutation}  D. Chevallier, D. Sticlet, P. Simon, C. Bena, Phys. Rev. B {\bf 85}, 235307 (2012).

\bibitem{chevallier_long}  D. Chevallier, D. Sticlet, P. Simon, C. Bena, Phys. Rev. B {\bf 87}, 165414 (2013).

\bibitem{dynes} R. C. Dynes, V. Narayanamurti, and J. P. Garno, Phys. Rev. Lett. {\bf 41}, 1509 (1978).

\bibitem{levy_yeyati} J. C. Cuevas, A. Martin-Rodero, and A. Levy Yeyati, Phys. Rev. B {\bf 54}, 7366 (1996).

\bibitem{chevallier_double_dot} D. Chevallier, J. Rech, T. Jonckheere, and T. Martin, Phys. Rev. B {\bf 83}, 125421 (2011).

\bibitem{jonckheere} T. Jonckheere,  G. Japaridze, T. Martin, R. Hayn, Phys. Rev. B {\bf 81}, 165443 (2010).

\bibitem{abrikosov} A. A. Abrikosov and L. P. Gor'kov, Zh. Eskp. Teor. Fiz. {\bf 39}, 1781 (1960)[Sov. Phys. JETP {\bf 12}, 1243 (1961)].

\bibitem{usadel} K. D. Usadel, Phys. Rev. Lett. {\bf 25}, 507 (1970).

\bibitem{houzet} Ya. V. Fominov, M. Houzet, and L. I. Glazman, Phys. Rev. B {\bf 84}, 224517 (2011).

\bibitem{usadel-mi} D. Chevallier,  C. Bena, and P. Simon, in preparation.

\end{thebibliography}
\end{document}